\documentclass[a4paper,10pt]{article}

\newcommand{\eqsection}{\makeatletter
    \@addtoreset{equation}{section}
    \renewcommand{\theequation}{\arabic{section}.\arabic{equation}}
    \makeatother}

\def\lal{&&\nqq {}}

\def\beq{\begin{equation}}
\def\eeq{\end{equation}}
\def\bear{\begin{eqnarray}}
\def\bearr{\begin{eqnarray} \lal}
\def\ear{\end{eqnarray}}
\def\earn{\nonumber \end{eqnarray}}

\def\dst{\displaystyle}
\def\tst{\textstyle}
\def\fracd#1#2{{\dst\frac{#1}{#2}}}
\def\fract#1#2{{\tst\frac{#1}{#2}}}
\def\Half{{\fracd{1}{2}}}
\def\half{{\fract{1}{2}}}

\title{An emergent universe supported by a Nonlinear Sigma Model}
\author{A. Beesham\thanks{E-mail: abeesham@pan.uzulu.ac.za} \\
 Department of Mathematical Sciences, Zululand University \\
 Private Bag X1001\\
Kwa-Dlangezwa 3886, South Africa \\S. V. Chervon\thanks{E-mail:
chervons@ukzn.ac.za
(On leave from Department of Theoretical Physics, Ulyanovsk State
University, Ulyanovsk, 432700 Russia)}
and S. D. Maharaj\thanks{E-mail: maharaj@ukzn.ac.za} \\
 Astrophysics and Cosmology Research Unit \\
School of Mathematical Sciences, University of KwaZulu-Natal \\
Private Bag X54 001 \\Durban 4000, South Africa \\}

\begin{document}

\maketitle

\begin{abstract}
We suggest the use of a nonlinear sigma model as the source which
supports an emergent universe. The two-component nonlinear sigma
model is considered as the simplest model containing inflaton and
auxiliary chiral fields.

\end{abstract}

\section{Introduction}

The emergent universe (EmU) scenario was originally proposed by
Ellis and Maartens in 2002 \cite{ellmaa02} with the aim of
avoiding the Big Bang singularity, and to open up the question
about the existence of quantum gravity. The EmU scenario has been
accepted by many cosmologists \cite{elmuts03, mukherjee05,
mukherjee06} as a viable model, and has been extended in many
aspects. Some of the recent treatments include the scalar-tensor
theory of gravity \cite{cahela07} and brane world gravity
\cite{babach07}.

In the original EmU model, Ellis and Maartens suggested the
existence of ordinary matter with energy density $\rho $ and
pressure $p=w\rho $. It is difficult to imagine such matter when $t
\rightarrow -\infty $, because of the very small volume of the
universe with the radius $a_i$, which is not very much larger than
the Plank scale. Therefore in the present article we suggest that
this ``ordinary matter'' be replaced with the chiral fields in the
framework of the non-linear sigma model (NSM). Chiral NSM were
introduced as a theory of strong interactions at the end of the
fifties by Schwinger \cite{schwinger57} and Skyrme \cite{skyrme58}.
Gell-Mann and Levy in their work \cite{gellev60} pointed out how to
realize the chiral symmetry and partial conservation of the axial
vector current. This article is often referred as the the work where
the terms ``linear and nonlinear sigma model'' have been introduced.
The investigation of the mathematical aspects of the NSM was closely
connected with the two dimensional version of the model because of
their analogy in many respects to non-Abelian gauge theories. The
main results of this investigations, such as soliton, instanton and
meron solutions, as well as the application of the inverse
scattering method application are summed up in the review
\cite{perelomov87}.

The consistent construction of the four dimension NSM can be
realized only by including a coupling to the gravitational field, as
was found when instanton solutions were investigated \cite{aff79}.
Consideration of NSM as the source of the gravitational field (with
Lorenz signature) was proposed by G. Ivanov \cite{ivanov83tmf} (see
also \cite{ch83iv}). The applications of the chiral NSM in general
relativity and cosmology have been examined also in
\cite{ch95gc,ch97gc}. In particular, it was shown that chiral NSM
with a potential of self-interaction (called in the framework of
cosmology -- {\it chiral cosmological model}) contains the
self-interacting scalar field (SSF) theories as well as
multicomponent ones. Therefore the chiral NSM can be considered as
an effective model and can describe not only SSF theory
\cite{ch94a}, but some classes of the
gravitational field as well \cite{matmis67}. 

Thus we can see the advantages and perspectives for the application
of the chiral NSMs to the EmU because of its wide physical content
and the fact that various geometrical methods can be applied in its
investigation.

Now we can apply the NSM as the source which support the EmU from
the very beginning (in description of negative times) to late times.
We shall start from the two component NSM as the simplest
possibility to find the solutions describing the features of the
model. The first scalar field we will consider is the inflaton while
an additional chiral field in this case may have various physical
interpretations. It may be the field inspired by superstring
cosmology \cite{lwc99}, such as the dilaton or moduli fields, or it
may be the scalar field responsible for dark matter or dark energy
\cite{mature00cqg}.
After studying the model with two fields, it will then be possible
to introduce more then one auxiliary chiral field with the aim of
considering interactions between dark matter, dark energy and the
inflaton as the source of the EmU.

There are a few approaches that we could follow in investigating
issues relevant in cosmology for NSM and SSF theory. The first
approach is the direct calculation with a given potential
\cite{ch97gc}.  In this approach we believe that high energy physics
(HEP) may provide us with the form of the potential $V$ of
self-interaction for the scalar field as the function on $\phi :
V=V(\phi) $. Then this dependence closes the set of Einstein and
scalar field equations, and we have to solve two equations with two
unknown functions: scalar field and the scalar factor as the
functions of cosmic time. The second approach which is also
connected with HEP is the fine turning method with a given scalar
field evolution $\phi=\phi (t)$ \cite{bar94}. In this approach the
dependence of scalar field on cosmic time is considered as given.
This fact also closes the set of equations and we can find, from
Einstein and scalar field equations, the scalar factor and the
potential of self-interaction as the function of cosmic time. The
third approach (the fine tuning of the potential method) is based on
astrophysical observations of the evolution of the scalar factor. We
can put as a physical reality the evolution of scalar factor
$a=a(t)$ into the set of Einstein and scalar field  equations. Then
we can find the potential $V(t)$ and scalar field evolution $\phi
=\phi (t)$. This method has been proposed in \cite{ellmad91}  and
developed in \cite{chzhsh97}, \cite{zhchsh98}. The investigation in
EmU is connected with the third approach, when the evolution of the
scalar factor is given {\it a priori}. Considering the fine tuning
of the potential method we can investigate matter which is
responsible for the evolution of the Universe that we observe. It is
very important to understand what types of matter manage the
expansion of the Universe and lie beyond this expansion. It is
almost evident that it may be various types of matter (equations of
state) that support observable expansion.  The fact that NSM can
provide support for the same gravitational field, as SSF theory, has
been proved earlier \cite{ch83iv}. Moreover it was found that the
effective scalar field represents NSM in the Einstein equations
\cite{ch97gc}. In this case we have to solve the dynamic equations
for the chiral fields. This result means that a single effective
scalar field may contain a great number of chiral fields, which can
represent dark matter, dark energy, phantom and quintessence fields,
etc. These fields, hidden in one effective scalar field, can exhibit
themselves in dynamical equations. Therefore it is of great
importance to study the chiral field equations because their
solutions will give new light for the better understanding of the
physical nature of hidden fields. There are other special methods of
exact cosmological solution construction in SSF theory such as
generation of new solutions with fine turned potential  $V=V(t)$
\cite{zhchsh98}, comparative analysis of slow-roll and exact
solutions in inflationary models \cite{zhch00} which lie outside our
consideration here.

The article is organized as follow. First, we introduce the main
equations of a self-gravitating NSM. In the next section we consider
the chiral cosmological model for the two component NSM. The
dynamical evolution of the chiral fields around the minimum of the
scale factor is considered for the first time in the EmU. We present
an example of the solution and investigate its asymptotic behaviour
when $t \rightarrow -\infty $. In Section 5, we investigate the
evolution of the chiral fields at late times. The two exact
solutions that we obtain involve new types of potential which
support the EmU at late times. We discuss our results in the last
section.


\section{A self-gravitating NSM}

The action for a self-gravitating NSM endowed with a
self-interacting potential $W(\varphi)$ is given by \cite{ch95gc,
ch01mg}

\beq\label{act1} {\cal S}= \int \sqrt{-g} d^4x \left\{
\frac{R}{2\kappa}+ \frac{1}{2}
 h_{AB}(\varphi) \varphi^A_{,\mu} \varphi^B_{,\nu} g^{\mu\nu} - W(\varphi) \right\}
\eeq
 Here $g_{\mu\nu}(x)$ is the space-time metric, and
 $h_{AB}(\varphi)$
 is the metric of the target space, $\varphi = (\varphi^{1},...,\varphi^{n}),
~\partial_\mu \varphi^A=\varphi^A_{,\mu}.$

For the chiral cosmological model, the energy-momentum tensor
corresponding to the action (\ref{act1}) reads
\beq\label{ch-em} T_{\mu \nu}=\varphi_{A_,\mu}\varphi^A_{,\nu}-
g_{\mu\nu}\Bigl\{\half\varphi^A_{,\alpha}\varphi^B_{,\beta}g^{\alpha\beta}
h_{AB}-W(\varphi^C)\Bigr\}.
\eeq Einstein's equations can be transformed to
 \beq\label{ein}
R_{\mu\nu}=\kappa\{ h_{AB}\varphi^A_{,\mu}\varphi^B_{,\nu}- g_{\mu
\nu} W(\varphi^C)\}.
 \eeq
 By varying the action (\ref{act1}) with
respect to the chiral fields $\varphi^C$, we can obtain the
equations of motion of the chiral fields as
 \beq\label{cfe}
\frac{1}{\sqrt{-g}}\partial_{\mu}\left(\sqrt{-g} \varphi^{,\mu}_A
\right) - \frac{1}{2} \frac{\partial
h_{BC}}{\partial\varphi^A}\varphi_{,\mu}^C\varphi_{,\nu}^B
g^{\mu\nu} + W_{,A} = 0,
 \eeq
where $W_{,A}=\frac{\partial W}{\partial \varphi^A}$.

\section{The chiral cosmological model}

If we will consider a self-gravitating NSM with the potential of
self-interaction in a homogeneous and isotropic universe, we arrive
at the  so-called \cite{ch01gcs} chiral cosmological model.

Now we consider the two-component nonlinear sigma model with the
diagonal target space metric
 \beq\label{tsmet} ds^2_{ts}=d\phi^2
+h_{22}(\phi,\psi)d\psi^2
 \eeq
In terms of the chosen target space (\ref{tsmet}), the
energy-momentum tensor (\ref{ch-em}) can be presented in the
following form
 \beq\label{em-2}
T_{\mu\nu}=\phi_{\mu}\phi_{\nu}+h_{22}\psi_{\mu}\psi_{\nu}-
g_{\mu\nu}\left[ \Half \phi_{\rho}\phi^{\rho}+ \Half
h_{22}\psi_{\rho}\psi^{\rho}- W(\varphi,\psi)\right]
 \eeq
Let us choose the metric of the homogeneous and isotropic universe
in the Friedmann-Robertson-Walker (FRW) form
 \beq\label{frw} ds^2=
dt^2-a(t)^2\left( \frac{dr^2}{1-K r^2} +r^2d\theta^2+r^2
\sin^2\theta d\varphi^2 \right)
 \eeq
The chiral field equations for the two-component NSM (\ref{tsmet})
in the FRW universe (\ref{frw}) can be represented as
\bear
\label{F1} \ddot\phi + 3H \dot\phi - \half \frac{\partial
h_{22}}{\partial \phi}\dot\psi^2
+\frac{\partial W}{\partial\phi}=0 \\
\label{F2} 3H\left( h_{22}\dot\psi \right) +\partial_t\left(
h_{22}\dot\psi \right) - \half \frac{\partial h_{22}}{\partial
\psi}\dot\psi^2 +\frac{\partial W}{\partial\psi}=0
 \ear
Einstein's equations can the be written in the form
 \bear\label{E1}
H^2=\frac{\kappa}{3}\left[\Half \dot\phi^2 +\Half h_{22}\dot\psi^2 +W \right]-\frac{K}{a^2} \\
\label{E2} \dot H =-\kappa \left[ \Half \dot\phi^2 +\Half
h_{22}\dot\psi^2\right]+\frac{K}{a^2}
 \ear
The overdot means a derivative with respect to the time variable
$t$. The Raychaudhuri field equation takes the form \beq
\frac{\ddot a}{a}=-\frac{\kappa}{3}\left[\dot\phi^2 +
h_{22}\dot\psi^2 -W \right] \eeq

From the Einstein and chiral field equations of the two-component
NSM, we notice that all features of the EmU scenario will be present
if the potential of $W(\phi,\psi)$ will contain the emergent
potential $V(\phi)$ \cite{elmuts03}, for example $W(\phi,\psi)=
V_{EmU}(\phi)+ \tilde{W}(\psi)$.

Let us consider now the two limiting cases, namely when $t
\rightarrow -\infty $ and when $t \rightarrow \infty $.

\section{Dynamics around the minimum}

Following the EmU scenario, let us consider the universe near the
minimum $a_{min}=a_i \equiv a(t_i)$, where $t_i \rightarrow
-\infty $. The only difference we suggest is that $ \ddot a \neq
0, \ddot a > 0$. This presentation of the dynamics in a small
local region, when time is close to $t_i$, provides the
possibility of studying the behaviour of the chiral fields
$\phi,\psi$ and the target space metric coefficients $h_{22}$ in
the EmU scenario.

The Einstein equations will take the following form (when
$a=a_i=const, \dot a=0,\ddot a > 0$)
\bear\label{E1-i}
\Half \dot\phi^2 +\Half h_{22}\dot\psi^2 +W =\frac{3K}{\kappa a^2} \\
\label{E2-i} \frac{\ddot a}{a}\mid_i =-\kappa \left[ \Half
\dot\phi^2 +\Half h_{22}\dot\psi^2\right]+\frac{K}{a^2}
\ear
Here the index ``i'' is omitted.

The chiral field equations take the following form
 \bear\label{F1-i}
\ddot\phi - \half \frac{\partial h_{22}}{\partial \phi}\dot\psi^2
+\frac{\partial W}{\partial\phi}=0 \\
\label{F2-i}
\partial_t\left( h_{22}\dot\psi \right) - \half
\frac{\partial h_{22}}{\partial \psi}\dot\psi^2 +\frac{\partial
W}{\partial\psi}=0 \ear For the scale factor, we take the form
\cite{mukherjee06}

 \beq\label{a-emu}
a(t)= a_i(\beta + e^{\alpha t}).
 \eeq

Our first task will be to find an example of the solution with any
potential which will support the stage under consideration. With
this aim, let us simplify the equations (\ref{E1-i}-\ref{F2-i}). Let
$h_{22}$ be functions of $\phi $ only. If $h_{22}$ will be an even
function of $\phi $ then the equations (\ref{E1-i}-\ref{F2-i})
possess the symmetry $\phi \leftrightarrow - \phi, ~\psi
\leftrightarrow -\psi $. Therefore we will keep only the positive
sign for the chiral fields. The resulting solution obtained under
the above assumptions can be presented by the formulae:
\bear\label{sol-phi}
\phi=\sqrt{\frac{2\lambda}{\kappa}}\int \sqrt{a_i^{-2}-\alpha^2
e^{\alpha t}}dt = \sqrt{\frac{2\lambda}{\kappa}}\left(2F(t,\alpha) +A^2\ln \left[\frac{F-A^2}{F+A^2}\right]\right)\\
\nonumber
~~\lambda= const,~~A^2=(a_i \alpha)^{-2},~~F(t,\alpha)=\sqrt{A^2-\exp (\alpha t)}\\
\label{sol-psi}
\psi = C_1 t,~~C_1=const\\
\label{sol-h}
h_{22}=\frac{2(1-\lambda)}{\kappa C_1^2}\left( a_i^{-2}-\alpha^2
e^{\alpha t}\right)=\frac{2\alpha^2(1-\lambda)}{\kappa C_1^2}F^2(t,\alpha)\\
\label{sol-w}
W= \frac{1}{\kappa}\left[ \alpha^2 e^{\alpha
t}+\frac{2}{a_i^2}\right]=W_1(\phi)+W_2(\psi)+W_*,
\ear
where \bear\label{W2}
W_2(\psi)=\frac{2\alpha^2}{\kappa}(1-\lambda)e^{\alpha\psi/C_1},\\
\label{W1} W_1= \frac{\alpha^2}{\kappa}(2\lambda-1)e^{\alpha t},\\
\label{W-0} W_*=\frac{2}{\kappa a_i^2}=const
\ear
It is clear that it is impossible to present the direct dependence
on $\phi $ for $h_{22}$ and $W_1$ in formulas (\ref{sol-h},
\ref{W1}). Asymptotically, when $t \rightarrow -\infty $, inflaton
$\phi \rightarrow -\infty,$ target space coefficient $
h_{22}\rightarrow \frac{2(1-\lambda)}{\kappa a_i^2 C_1^2},$ the part
of the potential $ W_1 \rightarrow 0.$ It needs to be remembered
that all obtained functions concern the period around $t_i$ and
formally should be provided by index "i".

Thus we showed that the asymptotically Einstein static solution in
the infinite past can be supported by the two chiral fields with
geometric interaction by means of NSM. The differences with the
original model of Ellis and Maartens are the following. If we
consider the first chiral field $\phi $ as the inflaton and the
$W_1$ part of the potential $W $ as the inflaton potential then we
conclude that the potential started from zero. This is directly
opposite to the original EmU scenario. But the presence of the
second chiral field $\psi $ may give another way of comparison our
EmU model with the original one. Namely, we have the possibility to
choose the sign of integration constant $C_1$. If we set $C_1<0$
then the potential $W_2(\psi)$ tends to infinity with $\psi
\rightarrow -\infty $ as in the original model with inflaton. In
this case we can redefine the second field as the inflaton. In this
situation another field will be responsible for geometric
interactions because this field is included in the metric of the
target space $h_{22}$.

The choice of integration constant $C_1>0$ will represent a new
situation in the infinite past. As we have already mentioned  for
the obtained solution (18)-(24):
$$ \phi \rightarrow
-\infty,~~W_1(\phi)\rightarrow 0,~~\psi \rightarrow
-\infty,~~W_2(\psi)\rightarrow 0,~~h_{22}\rightarrow
\frac{2(1-\lambda)}{\kappa a_i^2 C_1^2}
$$
This means that the geometrical interaction with constant value of
target space metric coefficient $h_{22}$ will support the Einstein
static regime instead of the potential with enormous large value at
the infinite past. Note that we are keeping $\ddot{a}_i \neq 0$
which was not considered in the original model. For the solution
presented we have
$$\ddot{a}_i=a_i \alpha^2 \exp (\alpha t)>0.$$

\section{Evolution at late times}

To consider the EmU model at late times, we will solve the
self-consistent system of the Einstein and chiral field equations
(\ref{F1}-\ref{E2}) with the scale factor
 \beq\label{a-lt}
 a(t)=\beta\exp{\alpha t}
\eeq
 An investigation of this epoch is also connected with our
attempts to understand the behaviour of the target space metric
coefficient $h_{22}$ and the potential of self-interaction
$W(\phi,\psi)$.

\subsection{Solution A}

The solution under the assumption $h_{22}=h_{22}(\phi)$ is the
following:
 \bear\label{lt-phi}
\phi=-\frac{1}{\alpha\beta}\sqrt{\frac{2(1-\lambda)}{\kappa}} e^{-\alpha t},\\
\label{lt-psi}
\psi = C_2 t,~~C_2=const\\
\label{lt-h}
h_{22}=\frac{2\lambda}{\kappa \beta^2 C_2^2}e^{-2\alpha t}=\frac{\lambda\alpha^2}{C_2^2(1-\lambda)}\phi^2\\
\label{lt-w} W= \frac{1}{\kappa}\left[ 3\alpha^2
+\frac{2}{\beta^2}e^{-2\alpha t}\right]=W_1(\phi)+W_2(\psi)+W_*,
\ear where \bear\label{lt-W2}
W_2(\psi)=\frac{\lambda}{\beta^2\kappa}e^{-2\alpha\psi/C_2},\\
\label{lt-W1} W_1= \frac{\alpha^2(2-\lambda)}{2(1-\lambda)}\phi^2,\\
\label{lt-W-0} W_*=\frac{3\alpha^2}{\kappa}=const
 \ear
 In this case we obtain the direct formulae for the potential
and for $h_{22}$. The dependence of $W$ on $\phi $ provides a
possibility to consider a very physical situation, viz., a massive
scalar field with the mass
$m_{infl}^2=\frac{\alpha^2(2-\lambda)}{(1-\lambda)}$ which supports
the epoch of late times in the EmU. It is clear from (\ref{lt-psi})
and (\ref{lt-W2}) that the influence of the second field $\psi $
will not be important when $t\rightarrow \infty $.

\subsection{Solution B}

We can obtain another solution under the same assumption for
$h_{22}: h_{22}=h_{22}(\phi)$. The solution is the following:
\bear\label{B-lt-phi}
\phi=-\frac{1}{\alpha\beta}\sqrt{\frac{2(1-\lambda)}{\kappa}} e^{-\alpha t},\\
\label{B-lt-psi}
\psi = -\frac{1}{s}e^{-st},~~s=const, ~~s\neq 0,~~s\neq\alpha\\
\label{B-lt-h} h_{22}=\frac{2\lambda}{\kappa \beta^2} e^{2(s-\alpha)
t}=
\frac{2\lambda}{\kappa\beta^2}B_*\phi^{\frac{2(\alpha-s)}{\alpha}}\\
\label{B-lt-w} W= \frac{1}{\kappa}\left[ 3\alpha^2
+\frac{2}{\beta^2}e^{-2\alpha t}\right]=W_1(\phi)+W_2(\psi)+W_*,
\ear
 where
\bear\label{B-lt-W2}
W_2(\psi)=\frac{\lambda s(s+\alpha)}{\alpha\beta^2\kappa}(-s)^{2\alpha/s-1}\psi^{2\alpha/s},\\
\label{B-lt-W1} W_1= \frac{\alpha}{2(1-\lambda)}\left\{2\alpha-\lambda(\alpha+s)\right\}\phi^2,\\
\label{B-lt-W-0} W_*=\frac{3\alpha^2}{\kappa}=const \ear
 The
constant $B_*$ in (\ref{B-lt-h}) can be written as
$$
B_*=\left[-\sqrt{\frac{2(1-\lambda)}{\kappa}}\frac{1}{\alpha\beta}\right]^{2(s-\alpha)/\alpha}
$$
The evolution of the scalar field $\phi $ with time is the same as
for the {\bf solution A}, but the ``effective'' mass of this field
is given by
$$
m_{infl}^2=\frac{\alpha}{(1-\lambda)}\left\{2\alpha-\lambda(\alpha+s)\right\}
$$
The second field $\psi $ also acquires an "effective" mass, but with
a dependence on $\psi $, since $s\neq \alpha $.

\section{Conclusions}

We have studied the Emergent Universe scenario within the context
of a nonlinear sigma model, and have shown that all the essential
features of the Emergent Universe are present in our case. Since
we know very little about the equation of state of ordinary matter
under extreme conditions, our description provides an alternative
scenario.

For the first time we investigated the dynamics of the source of the
EmU around the minimum of the scale factor when $t \rightarrow
-\infty $. The potential in this asymptotic case is proportional to
the exponent of both the fields $\phi $ and $\psi $.

Analysis of the evolution of the chiral field at late times gives a
new result for the potential form. Our two exact solutions show that
the evolution of the EmU at late times can be supported by a massive
inflaton field, while the auxiliary chiral field has exponential or
power law dependence. This form of the potential tells us about the
presence of the massive inflaton with some additional field during
the inflationary stage in the EmU. Thus the obtained solutions for
two limiting cases provide us some physical content about the chiral
fields and their interaction by means of the kinetic term $h_{22}$.

We have obtained two examples of exact solutions for late times with
a de Sitter expansion phase. Therefore we can consider the time of
exit from inflation $t=t_{end}$ as the limiting case for the
solutions. This time gives us restrictions on the parameters of the
solutions. As future work we wish to extend the exact inflation
method \cite{chefom08} for obtaining the key cosmological parameters
the for nonlinear sigma model. this will enable us tio compare
features from the presented solutions A and B with observation data.

\section{Acknowledgments}

SC is grateful to the University of KwaZulu-Natal, the University of
Zululand and the National Institute for Theoretical Physics of South
Africa for financial assistance and hospitality during his visit to
these universities. SC is partly supported by the Russian Foundation
for Basic Research (Project N 08-02-91307). SDM acknowledges that
this work is based on research by the South African Research Chair
Initiative of the Department of Science and Technology and the
National Research Foundation.

\end{document}